\documentclass[12pt, draftclsnofoot, onecolumn]{IEEEtran}
\usepackage{hyperref}
\usepackage{amsmath,amssymb}
\usepackage{latexsym}
\usepackage{CJK}
\usepackage{cite}   
\usepackage{caption} 

\usepackage{verbatim}  
\usepackage{graphicx}  
\usepackage{bm}  
\usepackage{mathrsfs} 
\usepackage{algorithm} 
\usepackage{algorithmic} 
\usepackage{booktabs}
\usepackage{subfigure}  
\usepackage{textcomp}  
\usepackage{multirow}  
\usepackage{lettrine}   
\usepackage{graphicx}  
\usepackage{color}  
\usepackage{amsmath}
\usepackage{amssymb}

\usepackage{algorithm}
\usepackage{algorithmic}

\usepackage{amsthm} 

\newtheorem{prop}{Proposition}
\newtheorem{rem}{Remark}

\newcommand{\myvec}[1]%
   {\stackrel{\raisebox{-2pt}[0pt][0pt]{\small$\rightharpoonup$}}{#1}}

\begin{document}
%
\title{Low-complexity Location-aware Multi-user Massive MIMO Beamforming for High Speed Train Communications}


\author{Xuhong Chen, Pingyi~Fan~\IEEEmembership{Senior Member,~IEEE,}\\
        State Key Laboratory on Microwave and Digital Communications, \\
        Tsinghua National Laboratory for Information Science and Technology, \\
        Department of Electronic Engineering, Tsinghua University, Beijing, China.\\
\thanks{This paper has been accepted for future publication by VTC2017-Spring}
E-mail: chenxh13@mails.tsinghua.edu.cn, fpy@mail.tsinghua.edu.cn}

\maketitle

\begin{abstract}
Massive Multiple-input Multiple-output (MIMO) adaption is one of the primary evolving objectives for the next generation high speed train (HST) communication system. In this paper, we consider how to design an efficient low-complexity location-aware beamforming for the multi-user (MU) massive MIMO system in HST scenario. We first put forward a low-complexity beamforming based on location information, where multiple users are considered. Then, without considering inter-beam interference, a closed-form solution to maximize the total service competence of base station (BS) is proposed in this MU HST scenario. Finally, we present a location-aid searching-based suboptimal solution to eliminate the inter-beam interference and maximize the BS service competence. Various simulations are given to exhibit the advantages of our proposed massive MIMO beamforming method.
\end{abstract}


\IEEEpeerreviewmaketitle

\section{Introduction}
\lettrine[lines=2]{W}ireless communication for high speed train (HST) has attracted much more attentions in recent years since billions of people tend to travel with HST and there is an ever-growing wireless communication demands for those passengers onboard. A real-scenario estimation \cite{demand} indicates that the demand could be as high as 65 Mbps with bandwidth 10 MHz for a train with 16 carriages and 1000 seats. To meet such huge demands, several innovative methods \cite{work, fan1, work4, work1} have been proposed to improve the information rate and system stability of wireless transmission. However, if the demands keep grow, promising techniques like massive Multiple-input Multiple-output (MIMO) and beamforming can be the potential candidates in the design of HST wireless communication systems.

In the literature, several preliminary attempts on integrate massive MIMO system or beamforming in HST wireless communication systems have been proposed. In \cite{work6}, a receive beamforming scheme has been proposed to reinforce the signal during the handover process in high mobility scenario. In \cite{work2} and \cite{work3}, a positioning-aided beamforming and gradual beamforming on handover scheme have been put forward for the HST wireless communication, respectively. An adaptive multi-stream beamforming for massive MIMO HST system aiming to improve the throughput has been proposed in \cite{work7}. Other related works have been listed in \cite{gain, work5}. However, to the best knowledge of us, low-complexity beamforming design and inter-beam interference elimination scheme for multi-user (MU) massive MIMO HST wireless communication system is still under-developed.

\begin{figure}[!b]
\centering
\includegraphics[width=0.48\textwidth]{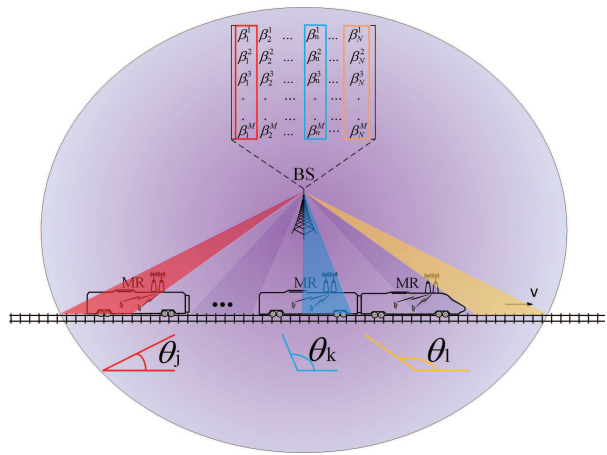}
\caption{The location-aided massive MIMO beamforming for HST scenario.} \label{fig:bf1}
\captionsetup{belowskip=-10pt}
\end{figure}
As to the HST scenario, there are several major challenges compared with conventional low mobility scenario. As illustrated in Fig. \ref{fig:bf1}, the train will suffer from large path loss when it traverses the coverage area of one base station (BS), which makes the large-scale fading dominant in this scenario compared with small-scale fading. Meanwhile, this dramatic fluctuation of received signal-to-noise ratio (SNR) at the mobile relay (MR) will challenge the conventional adaptive beamforming \cite{abf} or orthogonal switched beamforming \cite{obf} in channel detections. Besides, the channel detection process will be much harder not only because it appears fast time-varying and double-selective fading in spatial-temporal domains, but also because the wireless channel is hardly tractable (according to the estimation in \cite{demand}, the channel coherence time is less than 1 ms in HST scenario). More importantly, when massive MIMO is employed, the computational complexity needs to be seriously treated in beamforming design due to the aforementioned challenges, especially considering the limited service time for one BS.

However, there are certain advantages which can be exploited in this scenario to reduce the beamforming complexity. According to \cite{viaduct}, $86.5\%$ of the Chinese Beijing-Shanghai high speed railway is viaduct, which leads to dominant line-of-sight (LOS) and less scatters. Besides, the motion curve is fixed and the moving trend of HST should not be changed within a short time due to safety guarantee, which means the train location can be trackable and predictable. Actually in the HST scenario, the HST can only move along the rail, which indicates that no spatial-random burst communication requests will occur and therefore, narrows down the coverage scope of the beamforming scheme.

In this paper, we propose a location-aware low-complexity beamforming scheme for MU massive MIMO system in HST scenario. By exploiting the train location information, the beamforming can be performed without acquiring channel covariance matrix (CCM) or eigen-decomposing (ED) CMM. Consequently, the potential large complexity introduced by MU massive MIMO can be diminished. Since the service time of one BS is limited, we present a location-aided closed-form solution to maximize the total received mobile service of the whole train. Finally, based on the obtained closed-form resource allocation, we present a suboptimal searching-based algorithm to eliminate the inter-beam interference by utilizing the train location information.

The contributions of this paper can be concluded as
 \begin{itemize}
\item \emph{\textbf{A low-complexity beamforming for MU massive MIMO in HST scenario is presented.}}
\item \emph{\textbf{An optimal closed-form beamforming solution to maximize the BS total service competence is provided.}}
\item \emph{\textbf{A suboptimal location-aid searching-based algorithm to eliminate the inter-beam interference in this MU massive MIMO HST scenario is given.}}
\end{itemize}

The rest of this paper is organized as follows. Section II introduces the preliminary on system model, parameterizations and beamforming scheme. Section III presents the closed-form resource allocation solution to maximize the total received mobile service of the whole train without considering inter-beam interference. Section IV presents a suboptimal location adaptive searching-based solution to eliminate the inter-beam interference. Section V shows the numerical simulation results and the corresponding analyses. Finally, we conclude it in Section VI.

\section{Preliminary}

\subsection{System Model and Parameterizations}
Consider the downlink transmission of a MU massive MIMO system with location-aided beamforming in HST scenario. A two-hop structure is adopted as shown in Fig. \ref{fig:bf1}, because it avoids the large penetration loss caused by the metal body of train, namely, the BS transmission will be forwarded to users inside the train by the MR mounted on the top of the train. Thus, the total number of multi-user is equal to the number of carriages ($K$) in this scenario. The BS is located at the center of the cell and is equipped with an $M$-element uniform linear single-array antenna to form $N$ beams due to the \emph{3dB} gain over double-array structure \cite{gain, antenna}. The antenna spacing $d$ between each element is equal, where in this paper it is assumed $d=\frac{\lambda}{2}$ and $\lambda$ denotes the carrier wavelength. To the link between BS and the MR, the Doppler effect in HST scenario will not be considered here since it can be accurately estimated and removed from the signal transmission part as shown in \cite{work1, fan}.

As to the channel modeling, both large-scale fading (denoted as $G(t)$) and small-scale fading (denoted as $h(t)$) are considered in HST scenario. For simplicity, we adopt a normalized free-space path loss model to depict the large-scale fading, which is a function of the distance $d(t)$ between BS and the MR
\begin{equation}\label{G}
G(t)=\sqrt{\frac{G_c}{d^\alpha(t)}},
\end{equation}
where $G_c$ is normalized that $G_c=1$ and $\alpha\in [2, 5]$ represents the path loss exponent.

\subsection{The Location-aided Beamforming}
As aforementioned, the LOS signal is dominant and the angular spread around the MR is relatively tight in HST scenario. Therefore, location information can be exploited as a valuable side information to reduce the complexity of massive MIMO beamforming \cite{directional}. According to the previous research on massive MIMO beamforming design for HST scenario \cite{work5}, we adopt a location-aided beamforming scheme to further our analyses hereinafter, since this structure is applicable in HST scenario with considerable performance and low computational complexity. The instantaneous location information can be acquired or estimated by Global Positioning Systems, accelerometer and monitoring sensors along the railway as well as the train velocity with entrance time \cite{work4}. To save space, the effect of positioning error on beamforming performance will not be discussed here, which will be given in the extended version.

The BS beamforming process is illustrated in Fig. \ref{fig:bf1}, where the selection of antenna phase excitation $\bm{\beta}_n$ ($\bm{\beta}_n=[\beta_n^1, \beta_n^2, ... , \beta_n^M]^T$) to generate directional $n$-th beam ($n=1,2, ...,N$) at BS is determined by the acquired train location information. In this paper, the instantaneous location information $\theta_k(t)$ of carriage $k$ ($k=1,2,...,K$) is expressed as
\begin{equation}\label{L}
\theta_k(t)=\arcsin\frac{d_k(t)}{h},
\end{equation}
where $h$ represents the vertical distance from BS to rail. If Butler method is adopted to generate beam, the corresponding directivity for $n$-th beam directed to $k$-th carriage can be expressed as
\begin{equation}\label{D}
D_i(\theta_k(t))=\frac{2(AF_n(\theta_k(t)))^2}{\int_{0}^{\pi}(AF_n(\phi))^2sin\phi d\phi},
\end{equation}
where $AF(\cdot)$ is the array factor given by
\begin{equation}\label{AF}
AF_n(\theta_k(t))=\frac{sin(0.5N\pi cos(\theta_k(t))-b_n\pi)}{0.5N\pi cos(\theta_k(t))-b_n\pi},
\end{equation}
and
\begin{equation}\label{b}
b_n=-\frac{N+1}{2}+n.
\end{equation}

Note that $\beta_n=b_n\pi$ stands for the actual phase excitation on each element.
\begin{rem}
According to the definition of directivity $D$, when the train moves toward BS, the directivity will increase; when the train leave BS, the directivity will decrese; when $\theta_k(t)=\pi/2$, the directivity is maximized.
\end{rem}

Therefore, the beamforming weight of $n$-th beam can be expressed as
\begin{equation}\label{weight}
w_n=f_nD_n(\theta_k(t)),
\end{equation}
where $f_n=\sum\nolimits_{m=1}^Mf_n(m)$ denotes the power allocation coefficient for the $n$th beam and $f_n(m)$ stands for the actual amplitude excitation on the $m$-th ($m=1,2,...,M$) element for the $n$-th beam. For most of cases, the power for $n$-th beam is equal distributed to each element.

\section{Resource Allocation for Location-aid Beamforming}\label{Sec:RLandIE}

We consider the resource allocation for multiple carriages without inter-beam interference first, where only one beam is allocated to a single carriage. According to the aforementioned beamforming scheme, the received power of carriage $k$ can be expressed as
\begin{equation}\label{P}
P_k(t)=\sum_{n=1}^{N}i_{k,n}\overline{P}w_{n}(t)G_k^2(t)h_k^2(t),
\end{equation}
where $\overline{P}$ represents the constrained total transmit power. $N$ represents the total generated beam number and in order to avoid inter-beam interference, the selected beam number $N_s$ for carriage $k$ is one. $i_{k,n}\in\{0,1\}$ stands for an index of beam selection, where $i_{k,n}=1$ means beam $n$ is selected for carriage $k$ for transmission, otherwise $i_{k,n}=0$. Consequently, the power allocated for user $k$ ($p_{k,n}$) when beam $n$ is selected ($i_{k,n}=1$) can be expressed as: $p_{k,n}(t)=\overline{P}\cdot f_{k,n}(t)$. The corresponding normalized achievable rate for carriage $k$ can be expressed as
\begin{equation}\label{R}
R_k(t)=log(1+\gamma_k(t)),
\end{equation}
where
\begin{equation}\label{snr}
\gamma_k(t)=\frac{P_k(t)}{\sigma_k^2(t)}
\end{equation}
represents the MR received signal-to-noise ratio (SNR) and $\sigma_k^2(t)$ represents the power of additive white Gaussian noise at user $k$.

Unlike conventional low-mobility users, users on a high speed train will not stay in one BS for a long time, where the train speed usually around $350$ km/h and the duration in one BS usually is less than 10 seconds according to different BS deployments. Consequently, the instantaneous achievable rate can not fully exhibit the service competence of one BS when applying massive MIMO beamforming. Therefore, based on the achievable rate, we define the mobile service to quantify the service competence as the integral of the instantaneous achievable rate $R(t)$ over certain time period
\begin{equation}\label{ms}
S_k=\int_0^{t_s}R_k(\tau)d\tau,
\end{equation}
where $t_s$ represents the total service time of one BS coverage. Based on the previous analyses, our resource allocation target is to maximize the total mobile service of the train with diverse power allocation coefficient $f_{k,n}$ to each beam, which can be expressed as
\begin{align*}\label{equ:ms}
\tag{11}&\mathop{max}_{f_{k,n}} \,\, \sum_{k=1}^{K}S_k\\
\tag{12}&s.t.\quad  \sum_{k=1}^{K}\sum_{n=1}^{N}f_{k,n}(t)=1, f_{k,n}(t)>0.
\end{align*}

Note that $f_{k,n}(t)=0$ means no power is distributed to the $n$-th beam, that is, the $n$-th beam is not selected for user $k$ at system time $t$. The closed-form solution of this optimization problem is given in \textbf{Proposition 1}.

\begin{prop}Under the condition that transmit CSI is not known, the optimal power allocation coefficient satisfies
\begin{equation*}\label{prop1}
f_{k,n}(t)=max\left\{0,\frac{1}{\lambda K\overline{P}}-\frac{1}{\overline{P}\sum_{n=1}^{N}i_{k,n}D_{n}(t)\mathcal{W}_k(t)}\right\} \tag{13}
\end{equation*}
and
\begin{equation*}\label{trans2}
\lambda=\frac{1}{\overline{P}+\sum\limits_{k=1}^{K}\frac{1}{\sum_{n=1}^{N}i_{k,n}D_n(t)\mathcal{W}_k(t)}}. \tag{14}
\end{equation*}
\end{prop}

\noindent\emph{Proof.} Let us consider the Lagrangian
\begin{equation*}
\begin{aligned}
F_w&=\int_{0}^{t_s}\left [\sum_{k=1}^{K}R_k(\tau)\right ]d\tau-\lambda(\int_{0}^{t_s}\sum_{k=1}^{K}\sum_{n=1}^{N}p_{k,n}(\tau)d\tau-\overline{P}t_s)\\
&=\int_{0}^{t_s}\left [\sum_{k=1}^{K}R_k(\tau)-\lambda\left (\sum_{k=1}^{K}\sum_{n=1}^{N}p_{k,n}(\tau)-\overline{P}\right )\right ]d\tau \\
&\triangleq\int_{0}^{t_s}L_w(\tau)d\tau
\end{aligned}
\end{equation*}
where $L_w(\tau)=\sum\limits_{k=1}^{K}R_k(\tau)-\lambda\left (\sum\limits_{k=1}^{K}\sum\limits_{n=1}^{N}p_{k,n}(\tau)-\overline{P}\right )$. Since Large-scale fading is dominant in this scenario, we assume $h(t)=1$ \cite{h1} and set $\mathcal{W}_k(t)=G_k^2(t)/\sigma_0^2$. Based on Euler’s formula, performing derivative of this integrand is equivalent to performing the derivative of $F_w(t)$. Therefore,
\begin{equation*}
\begin{aligned}
\frac{\partial L_w(t)}{\partial p_{k,n}(t)}&= \frac{\sum_{n=1}^{N}i_{k,n}\cdot D_n(t)\cdot \mathcal{W}_k(t)}{1+\sum_{n=1}^{N}i_{k,n}\cdot p_{k,n}(t)\cdot D_n(t)\cdot \mathcal{W}_k(t)}-\lambda K.
\end{aligned}
\end{equation*}

Setting $\partial L_w(t)/\partial p_{k,n}(t)=0$, we get
\begin{equation*}
p_{k,n}(t)=\frac{1}{\lambda K}-\frac{1}{\sum_{n=1}^{N}i_{k,n}\cdot D_n(t)\cdot \mathcal{W}_k(t)}.
\end{equation*}

Then the optimal power allocation coefficient is
\begin{equation*}
f_{k,n}(t)=\frac{1}{\lambda K \overline{P}}-\frac{1}{\overline{P}\sum_{n=1}^{N}i_{k,n}\cdot D_n(t)\cdot \mathcal{W}_k(t)}.
\end{equation*}

To find this $\lambda$, we utilize the power allocation coefficient constrain that $\sum_{k=1}^{K}\sum_{n=1}^{N}f_{k,n}(t)=1, f_{k,n}(t)> 0$. Then we have
\begin{equation*}
\sum\limits_{n=1}^{N}\left [\frac{1}{\lambda K \overline{P}}-\frac{1}{\overline{P}\sum_{n=1}^{N}i_{k,n}\cdot D_n(t)\cdot \mathcal{W}_k(t)}\right]=1.
\end{equation*}

By solving this equation, one can eventually obtain
\begin{equation*}
\lambda=\frac{1}{\overline{P}+\sum\limits_{k=1}^{K}\frac{1}{\sum_{n=1}^{N}i_{k,n}\cdot D_n(t)\cdot \mathcal{W}_k(t)}}.
\end{equation*}

This completes the proof of \textbf{Proposition 1}.

\begin{rem}
Based on \textbf{Remark 1} and \textbf{Proposition 1}, to maximum the total mobile service, the pre-set power allocation coefficient $f_{k,n}$ for the selected beam increases when one carriage moves toward to BS and decreases when it leaves BS.
\end{rem}

\section{Inter-beam Interference Elimination}

Based on the previous discussion, we consider a much complicated situation that during the beam selection process, more than one beam are selected for a single carriage ($N_s>1$), which may trigger inter-beam interference. For example, this situation may occur when the train location information is temporarily inaccurate and thus a wider beam (consider as the combination of the selected beams) is choosen. The inter-beam interference is defined as one beam is utilized by multiple users, where for most case in HST scenario, it stands for one beam is utilized by different carriages within the same BS. To be more specific, the inter-beam interference will occur with large probability between adjacent carriages under this LOS scenario.

Therefore, during the beam selection process, the received power of carriage $k$ can be expressed as
\begin{equation*}\label{P}
P_k^{'}(t)=\sum_{n=1}^{N}i_{k,n}\overline{P}w_n(t)G_k^2(t)h_k^2(t). \tag{15}
\end{equation*}

Consequently, the inter-beam interference for carriage $k$ can be expressed by
\begin{equation*}\label{interference}
I_k(t)=\sum_{j=1,j\ne k}^{K}\sum_{n=1}^{N}i_{j,n}\overline{P}w_n(t)G_j^2(t)h_j^2(t), \tag{16}
\end{equation*}
where $i_{j,n}$ stands for the index of interference beam and it subjects to $\sum_{j=1,j\ne k}^{K}\sum_{n=1}^{N}i_{j,n}\le \sum_{n=1}^{N}i_{k,n}$.

Therefore, the achievable rate can be expressed as
\begin{equation*}\label{R1}
R_k^{'}(t)=log(1+\Gamma(t)), \tag{17}
\end{equation*}
where
\begin{equation*}\label{sinr}
\Gamma(t)=\frac{P_k^{'}(t)}{I_k+\sigma_k^2(t)} \tag{18}
\end{equation*}
represents the received signal-to-interference-plus-noise ratio (SINR).

The object here is to maximize the total mobile service of the whole train when considering the inter-beam interference for location-aided massive MIMO beamforming, which can be expressed as the following optimization problem.
\begin{align*}\label{equ:ms1}
\tag{19}&\mathop{max}_{i_{k,n}} \,\, \sum_{k=1}^{K}\int_0^{t_s}R^{'}_k(\tau)d\tau\\
\tag{20}&s.t.\quad  \sum_{j=1,j\ne k}^{K}\sum_{n=1}^{N}i_{j,n}\le \sum_{n=1}^{N}i_{k,n}, i_{k,n}\in\{0,1\}\\
\end{align*}

Because the above optimization problem is non-convex, we aim to solve it with a sub-optimal searching-based algorithm instead of an optimal closed-form solution. For a high-speed train, even the carriage number of a train is limited, the computational complexity of full-search algorithm will probably leads to performance degradation in real-scenario application since massive MIMO is considered. Fortunately, according to the previous conclusions in \textbf{Remark 1} and \textbf{Remark 2}, we find that the beam near the center of BS possesses a higher directivity and also occupies a larger power allocation coefficient due to the better channel condition, which inspires us to optimize the searching algorithm based on the train location again, namely, if two adjacent carriages select one or several common beams in the same time, the carriage near the BS has the priority to occupy a beam with higher gain.

Therefore, based on the aforementioned discussion, the location-aided beam-selection searching algorithm is given in Algorithm 1.

\begin{algorithm}\label{alg:lsearch}
\caption{The location-aided beam selection against inter-beam interference}
\begin{algorithmic}
\STATE \textbf{Initialization}:
\STATE \quad $i_{k,n}$, $a[\,]=[d_1(t),d_2(t),...,d_K(t)].$
\STATE 1. \textbf{for} $k=1$ to $K$ \textbf{do}
\STATE 2. \quad \ $k^*=\mathop{\arg\min}\limits_{k}d_k(t)$;
\STATE 3. \quad \ \textbf{for} $n=1$ to $N$ \textbf{do}
\STATE 4. \quad \quad \quad \textbf{if} $i_{k^*,n}=1, k^*+1 \le K$ \textbf{then}
\STATE 5. \quad \quad \quad \quad  $n^*=n, i_{k^*+1,n}=0$;
\STATE 6. \quad \quad \quad \textbf{else if} $i_{k^*,n}=1, k^*-1 \ge 1$ \textbf{then}
\STATE 7. \quad \quad \quad \quad  $i_{k^*-1,n}=0$;
\STATE 8. \quad \quad \quad \textbf{end if}
\STATE 9. \quad \ \textbf{end for}
\STATE 10. \quad \ $a[k^*]=0$;
\STATE 11. \textbf{end for}
\end{algorithmic}
\end{algorithm}

Therefore, the whole beam scheduling process to avoid inter-beam interference is based on the continuous feed back of train location information. For instance, during the time the train traverses the BS coverage, before the phase excitations for the selected beams directed to each carriage are applied by the antenna, the closest carriage (i.e. the 2th carriage) will check whether the selected beams (i.e. the 7th and 8th beam) are also occupied by other carriages (i.e. the 1st or 3rd carriage). If occupied (the 1st carriage also select the 7th beam for transmission), other carriage will release the beams (7th beam) and re-select. After it, the second closest carriage will also perform the same operation until the last carriage is done.

\section{Numerical Results}\label{Sec:NumRes}
In this section, several numerical results are presented. Assume that $\alpha=3$, $h=50$m, $N=32$, carriage length $L_c=25$m and train velocity $v = 360$km/h.

\begin{figure}[t]
  \centering
  \subfigure[30dBm]{\label{fig:bf:a}
    \includegraphics[width=0.5\textwidth]{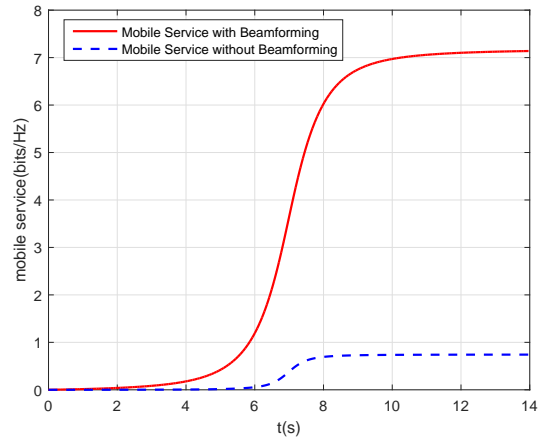}}
  \hspace{1in}
  \subfigure[40dBm]{\label{fig:bf:b}
    \includegraphics[width=0.5\textwidth]{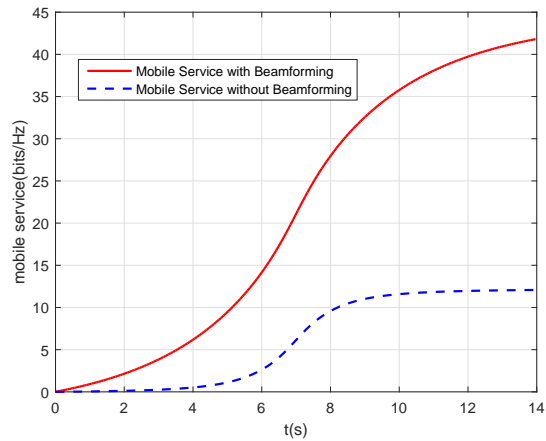}}
  \caption{Mobile service comparison under different transmit power with/without beamforming.}
  \label{fig:bf}
\end{figure}
In Fig. \ref{fig:bf}, a cumulative mobile service comparison under diverse transmit power is exhibited since it stands for the overall service competence of one BS, which in fact, is the maximum data volume that can be transmitted during the traversing time. A significant performance improvement can be achieved by adopting this location-aware beamforming scheme in this HSR scenario and it is noticed that the improvement is much obvious when the transmit power is lower, where equal power allocation policy is adopted for beamforming.

\begin{figure}[h]
  \centering
  \subfigure[30dBm]{\label{fig:subfig:a}
    \includegraphics[width=0.5\textwidth]{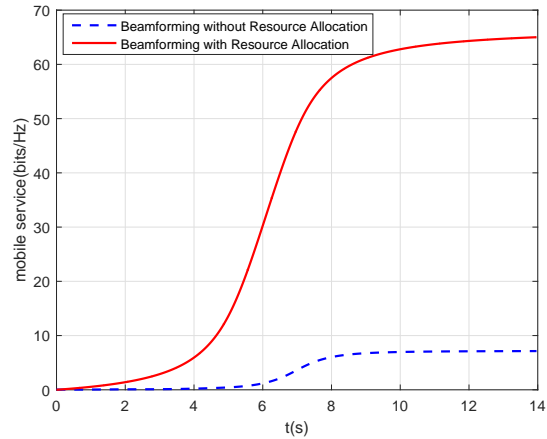}}
  \hspace{1in}
  \subfigure[40dBm]{\label{fig:subfig:b}
    \includegraphics[width=0.5\textwidth]{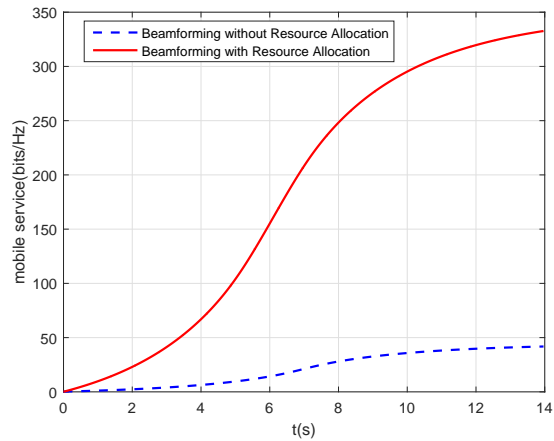}}
  \caption{Mobile service comparison under different transmit power with/without resource allocation.}
  \label{fig:subfig}
\end{figure}

As shown in Fig. \ref{fig:subfig}, the mobile service with different resource allocation schemes are illustrated, where beamforming without resource allocation stands for equal transmit power allocation for each selected beam. The simulation results indicate that even when the instant CSI is not available, location-aware resource allocation for each beam is crucial for better performance because large-scale fading is dominant in this scenario and the location information is known. Besides, the whole resource allocation process can be also pre-calculated as the beamforming weight, which reduces the system complexity as well.

\begin{figure}[t]
\centering
\includegraphics[width=0.48\textwidth]{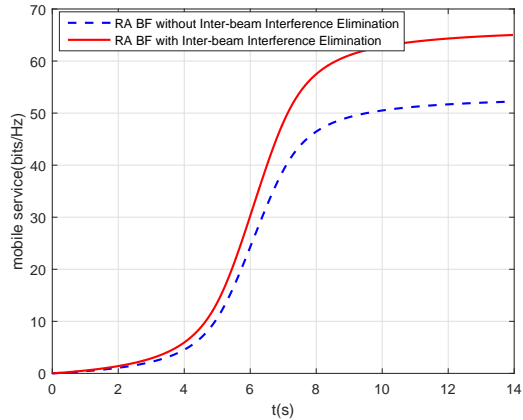}
\caption{Mobile service comparison with/without inter-beam interference elimination.} \label{Fig:bVersusThetab}
\end{figure}

In addition, Fig. \ref{Fig:bVersusThetab} shows that when multiple carriage occasion is considered and multiple beams are selected, inter-beam interference elimination is paramount in the beamforming application process, where this improvement will be much obvious when two trains encounter. By applying a location-aided interference elimination method, this performance degradation is eliminated.

\section{Conclusions}

In this paper, we first put forward a low-complexity beamforming of MU massive MIMO system for HST. By exploiting new side information in this LOS scenario, we managed to perform beamforming with the aid of trackable and predictable train information, which in the meantime diminishes the complexity of massive MIMO system. Then, to improve the service of the BS, we proposed a closed-form resource allocation scheme to maximize the whole service competence that one train can receive. Finally, inter-beam interference was considered in this beamforming scheme since multiple carriages are involved. By location-aided beam scheduling, a searching-based suboptimal algorithm was proposed to eliminate this interference.

\section*{Acknowledgement}

This work was supported by State Key Development Program of Basic Research of China No. 2012CB316100(2) and National Natural Science Foundation of China (NSFC) No. 61321061.
\end{document}